\documentclass[aps,prl,twocolumn,showpacs,amsmath,amssymb]{revtex4}
\usepackage{epsfig}
\usepackage{bm}
\def\be{\begin{equation}}
\def\ee{\end{equation}}

\begin{document}
\title{Stochastic Path Integral Formulation of Full Counting Statistics}
\author{S. Pilgram, A. N. Jordan, E. V. Sukhorukov and M. B\"uttiker}
\affiliation{D\'epartement de Physique Th\'eorique, Universit\'e de Gen\`eve,
        CH-1211 Gen\`eve 4, Switzerland}
\date{\today}

\begin{abstract}
We derive a stochastic path integral representation of counting statistics
in semi-classical systems.  
The formalism is introduced on the simple case of a single
chaotic cavity with two quantum point contacts, and then further generalized 
to find the propagator for charge distributions 
with an arbitrary number of
counting fields and generalized charges.
The counting statistics is given by the saddle point
approximation to the path integral, and fluctuations around the saddle
point are suppressed in the semi-classical approximation.
We use this approach to derive the 
current cumulants of a chaotic cavity in the hot-electron regime.  
\end{abstract}
\pacs{73.23.-b, 05.40.-a, 72.70.+m, 02.50.-r, 76.36.Kv}

\maketitle
Noise properties of electrical conductors are 
interesting because they reveal additional information
beyond linear response \cite{BB}.
In the pioneering work of Levitov and Lesovik \cite{Levitov1}, 
the optics concept of full counting statistics (FCS) 
for photons was introduced for electrons
in the context of mesoscopic physics.
FCS gives the probability of counting
a certain number of particles at a measurement 
apparatus in a certain amount of time and
finds not only conductance and shot noise, 
but all higher current cumulants as well.  
Several methods have been used in finding this
quantity.
Originally, quantum mechanical methods 
based on scattering theory
\cite{Levitov1,Muzykantskii1}, 
the Keldysh approach put forth by Nazarov \cite{Nazarov1}
or sigma-models \cite{Gefen1}
have been advanced and have been successfully applied 
to a number of problems among which we mention only 
multiterminal structures \cite{Bagrets1}, 
normal-superconducting
samples \cite{Belzig1}, 
combined photon/electron statistics \cite{Schomerus1},
and conductors which are current (instead of voltage)  
biased \cite{Kindermann1}.

A quantum mechanical treatment of transport shows that the
leading contribution to current cumulants is of the 
order of the channel number $N$. 
For many conductors or circuits of interest, 
this leading order is a semi-classical 
quantity \cite{Beenakker1}. 
Weak localization or universal conductance 
fluctuations provide only a small correction of order $1$.  
Clearly, it is desirable to have a purely semi-classical 
theory to calculate semi-classical results. 
To provide such a derivation of FCS 
is the main purpose of this work. 

\begin{figure}[htb]
\begin{center}
\leavevmode
\psfig{file=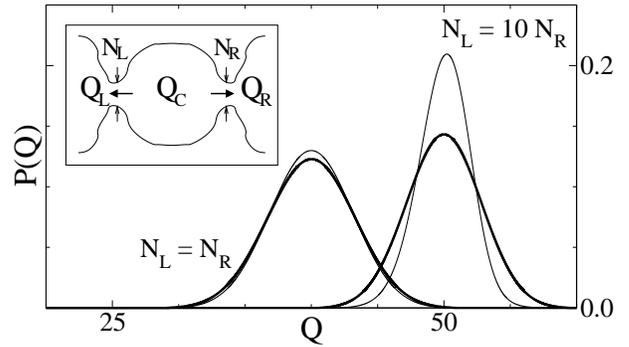,width=8cm}

\caption{Full counting statistics
of a chaotic cavity at zero temperature:
Comparison between hot (thick lines) and cold (thin lines) electron
regime for both symmetric ($\langle Q\rangle=40e$) and asymmetric 
($\langle Q\rangle=50e$) cases.
Inset: The chaotic cavity with two point contacts.
}
\label{Probability Distribution}
\end{center}
\vspace{-10mm}
\end{figure}

That a purely classical theory should be developed 
was realized by 
de Jong \cite{DeJong1} 
who put forth a discussion for problems which can be described 
with the help of master equations. A more general approach, 
leading to a set of rules for a cascade 
of higher order cumulants, was recently invented 
by Nagaev \cite{Nagaev1} and applied to chaotic cavities \cite{Nagaev2}. 
The work presented here aims at providing a foundation 
for the cascade approach by deriving a functional integral
from which FCS, but also dynamical quantities such as correlation functions,
can be obtained. 

The approach provided here applies to an arbitrary mesoscopic 
network. Its semi-classical nature 
does not allow the treatment of 
weak localization corrections nor is it applicable 
to macroscopic quantum effects like the proximity 
effect.
As in Boltzmann-Langevin theory \cite{Kogan1}, 
our approach is based on a separation of time scales: 
fast microscopic fluctuations causes variations of conserved 
quantities (like the charge inside the conductor) on much longer 
time scales. Whenever such a separation of time scales 
is present in a stochastic problem, the formalism outlined here 
is useful. This includes for example
normal-diffusive wires \cite{Nagaev1,Henny},  superconducting-normal 
structures outside the proximity regime \cite{Sanquer}, as well as 
stochastic problems beyond mesoscopic physics \cite{Langevin}. 

{\em Path integral derivation.}
To introduce our path integral formalism, we consider a simple 
example of electron transport through a chaotic cavity 
(see Fig.\ \ref{Probability Distribution}) which is a large conductor connected 
to two metallic leads $L$ and $R$ through
ideal point contacts. The leads are at different 
chemical potentials $\mu_R$ and $\mu_L=\mu_R+eV$, causing 
a current $I$ to flow through the cavity. 
The number of modes in each point contact $N_{L,R}$,
and the number of states $eVn_F$ ($n_F$ being a Fermi density of
states in the cavity) are large,
so that the transport is classical.
We assume elastic transport (no energy relaxation), 
zero temperature, and chaotic electron dynamics in the cavity
\cite{commentm1}. Then, the state of the cavity is described by
only one variable $f_C$ which is
independent of the electron's coordinate, momentum and energy
in the interval $\mu_R<\varepsilon<\mu_L$ \cite{Langen,cavity}.
The average values of outgoing currents $I_{L,R}$ 
are given by 
$\langle I_R\rangle=(e^2V/2\pi\hbar)N_R\langle f_C\rangle$ and 
$\langle I_L\rangle=-(e^2V/2\pi\hbar)N_L(1-\langle f_C\rangle)$. 
Since currents are conserved
$\langle I_L\rangle+\langle I_R\rangle=0$, the average current
through the cavity is $\langle I\rangle=
(e^2V/2\pi\hbar)[N_LN_R/(N_L+N_R)]$, i.e.\ the resistances
of the two point contacts add. 

The currents $I_{L,R}$ however fluctuate as a result of the discretness 
of the electron charge.
These fast fluctuations (correlated on the time scale $\tau_0\sim \hbar/eV$) 
cause slow variations of the occupation $f_C$ on the scale of the
relaxation time $\tau_C=2\pi\hbar n_F/(N_L+N_R)$ 
\cite{comment0,Brouwer}. The cause of this is charge conservation:   
$I_L+I_R=-\dot Q_C$, where $Q_C=e^2n_FVf_C$ is the charge accumulated in the cavity.
Fluctuations of $f_C$ in turn affect the currents $I_{L,R}$ thereby generating 
correlations between them. Our goal is to integrate out currents 
$I_{L,R}$ taking into account correlations and to obtain the FCS 
of the transmitted
charge $Q(t)=(1/2)\int^{t}_{0}dt' [I_R(t')-I_L(t')]$. More precisely, we  
have to find the generator $S(\chi,t)$ of the charge cumulants defined as
\begin{eqnarray}
P(Q,t)&=&(2\pi)^{-1}\int d\chi \exp[-i\chi Q+S(\chi,t)],
\label{P-def}\\
\langle Q^m(t)\rangle &=&\partial^m S(\chi,t)/\partial(i\chi)^m|_{\chi = 0}
\label{cumulants-def}
\end{eqnarray}
where $P(Q,t)$ is the distribution of transmitted charge,
and $\langle Q^m(t)\rangle$ is its $m$-th cumulant. 
The average current and its zero-frequency noise power are given respectively
by the first cumulant $\langle Q(t)\rangle/t$, and second 
cumulant $\langle Q^2(t)\rangle/t$, after taking the limit $t\to\infty$.

The separation of time scales 
$\tau_C/\tau_0\gg 1$ allows us to consider the intermediate time interval $\Delta t$,
such that $\tau_0\ll\Delta t\ll\tau_C$, on which the occupation $f_C$ is approximately
constant in time. Then the charges transmitted through the point contacts 
during this time interval, 
$Q_{L,R}=\int^{\Delta t}_0 dt' I_{L,R}$, are 
not correlated,  i.e.\ the total distribution is a product
$P_L(Q_L)P_R(Q_R)$. On the other hand, 
since $\Delta t$ is large compared to the correlation time $\tau_0$,
their distributions $P_{L,R}$ can be written as Fourier integrals (\ref{P-def})
with the cumulant generators proportional to time $\Delta t$ \cite{comment1},
namely $S_{L,R}=H_{L,R}\Delta t$.
They are well known \cite{Levitov1} and for ideal point
contacts given by
\begin{eqnarray}
\label{QPCGenerators}
H_{L,R}
&=&(N_{L,R}/2\pi\hbar)
\int d\varepsilon\left\{
\ln[1+f(\varepsilon)(e^{ie\chi_{L,R}}-1)]\right.\nonumber\\
&+&\left.\ln[1+f_{L,R}(\varepsilon)(e^{-ie\chi_{L,R}}-1)]\right\}.
\label{H-LR}
\end{eqnarray}
For elastic transport, only the energy interval 
$\mu_R<\varepsilon<\mu_L$ contributes with $f_L=1$, $f_R=0$,
and $f=f_C$. 

To proceed with our derivation we first discretize time, 
$t_n=n\Delta t$, $n=1,\ldots,N$,
and write the characteristic function of transmitted charge
$\exp[S(\chi,t)]$ for the unconditional probability distribution
\begin{eqnarray}
\exp[S(\chi,t)]&=&\prod_n\prod_{l=L,R}\int dQ_l(t_n)P_l[Q_l(t_n)]\nonumber\\
&\times &
\exp\bigg\{\frac{i\chi}{2}\sum\limits_n[Q_R(t_n)-Q_L(t_n)]\bigg\}.
\label{uncorr-S}
\end{eqnarray}
Next, we impose a constraint 
\begin{equation}
f_C(t_{n+1})=f_C(t_n)-(e^2n_FV)^{-1}[Q_L(t_n)+Q_R(t_n)],
\label{constraint}
\end{equation}
which guarantees the conservation of charge on the cavity, and
gives dynamics to the fluctuating charge $Q_C$.
Integrating out the charges $Q_{L,R}(t_n)$ in Eq.\ (\ref{uncorr-S})
with the use of Eqs.\ (\ref{P-def}) and (\ref{constraint}), and taking a continuum 
limit, we obtain one of our main results
\begin{eqnarray}
&&\exp[S(\chi,t)]
=(2\pi)^{-1}\int {\cal D}Q_C{\cal D}\chi_C\nonumber\\ 
&&\qquad\times\exp\left\{\int_0^t dt'[i\chi_C\dot Q_C
+H(\chi, \chi_C,Q_C)]\right\},
\label{corr-S}
\end{eqnarray}
where $H(\chi,\chi_C,Q_C)$ is 
\begin{equation}
H=H_L(\chi_C-\chi/2,f_C)+H_R(\chi_C+\chi/2,f_C)
\label{Hamiltonian}
\end{equation}
with $H_{L,R}$ given by Eq.\ (\ref{H-LR}).
Eq.\ (\ref{corr-S}) resembles the imaginary time
path integral for the evolution operator of a quantum particle
with coordinate $Q_C$, momentum $\chi_C$, and Hamiltonian $-H$
with the difference that $H$ given by (\ref{Hamiltonian})
is not Hermitian. Nevertheless, we can apply the Hamiltonian formalism
to our path integral (\ref{corr-S}). Next, we derive the saddle-point 
solution and discuss its applicability. 

{\em Saddle-point solution.}
The saddle point of the path integral (\ref{corr-S})
can be obtained by a variation of the action with respect to $\chi_C$ and
$Q_C$ and gives ``classical equations of motion''
\begin{equation}
i\dot Q_C=-\partial H/\partial\chi_C,\quad
i\dot\chi_C=\partial H/\partial Q_C.
\label{motion}
\end{equation}
These equations have to be solved with the conditions
$Q_C=Q_C(0)$ for $t'=0$, and $\chi_C=0$ for $t'=t$ \cite{comment2}.
The solution has to be substituted into Eq.\ (\ref{corr-S})
to obtain the saddle-point part $S_0$ of the cumulant generator.

Eqs.\ (\ref{motion}) describe 
the relaxation of the initial state $Q_C(0)$ to the stationary state
$\{\bar\chi_C,\bar Q_C\}$
given by the solution of $\partial H/\partial\chi_C=
\partial H/\partial Q_C=0$. On time scales large compared
to the relaxation time $\tau_C$, the initial and final integration points
contribute little (see discussion below), and we can neglect the first term 
in the action (\ref{corr-S}). This gives us a large time asymptotics
\cite{SP-solution},
\begin{equation} 
S_0(\chi, t)=tH(\chi,\bar\chi_C,\bar Q_C),\quad t\gg \tau_C,
\label{asymptotics}
\end{equation}
which will be analyzed below.

To further simplify the analysis, we concentrate on the 
symmetric cavity with equal number of modes $N_L=N_R=N_{pc}$
in the contacts. Then the stationary saddle-point 
solution can be found analytically giving
$\bar\chi_C=0$, $\bar f_C=1/2$, and 
\begin{equation} 
S_0(\chi, t)= (eVN_{pc}t/\pi\hbar)\ln[(1/2) 
(1+e^{ie\chi/2})]
\label{symmetric-S}
\end{equation}
which is the known result 
for a symmetric cavity found from quantum mechanical 
calculations \cite{Blanter}.

To investigate the validity of the saddle-point solutions (\ref{motion}) and 
(\ref{asymptotics}) we calculate the contribution of Gaussian fluctuations
by expanding $S$ in the vicinity of the 
stationary point to second order in $\chi_C$ and $Q_C$.
For the correction to $S_0$ we 
obtain the action of a harmonic oscillator 
in imaginary time \cite{comment3} 
with inverse frequency $\omega^{-1}=\tau_C\cos(e\chi/4)$
and mass $M=(4\pi\hbar/e^3VN_{pc})\cos^2(e\chi/4)$. In the context 
of our problem such an action describes the linear
dissipative dynamics  $\dot Q_C=-\omega (Q_C-\bar Q_C)+\nu(t)$ 
with the Gaussian Langevin source $\langle
\nu(t)\nu(t')\rangle=M^{-1}\delta(t-t')$
\cite{Langevin},
which takes the initial state $Q_C(0)$ to the stationary
state $\bar Q_C$ after time $t\sim\tau_C$. Integrating out $Q_C$ we obtain
for the total action $S=S_0+\delta S_0+S_{fl}$, where 
$\delta S_0=-(1/2)M\omega Q_C^2(0)\tanh(\omega t)$ is the initial 
condition contribution and $S_{fl}=-(1/2)\ln[\cosh(\omega t)]$
is the fluctuation contribution. Since in general $Q_C(0)\alt e^2n_FV$,
the part $\delta S_0$ is small compared to $S_0$ if $t\gg\tau_C$. 
The part $S_{fl}$ is small compared to $S_0$ if $en_FV\gg 1$, which has been 
assumed for our classical cavity \cite{comment4}. All this justifies 
the saddle-point solution (\ref{asymptotics}). Having established 
the framework of our formalism, we now proceed with its generalization and
applications.

{\em Generalization.}
We consider the nonequilibrium dynamics of an arbitrary classical system,
which can be described by fast fluctuating 
currents $I_{\alpha\beta}$
and slow fluctuating charges $Q_{\alpha}$. Metallic reservoirs, such as 
leads and cavities, can be taken into account by replacing
$\alpha\to(\alpha,l)$, where the index $l=1,\ldots,L$ enumerates reservoirs,
and $\alpha$ enumerates generalized charges in the reservoirs.
These charges $Q_{\alpha}$ are defined as
$\dot Q_{\alpha}=-\sum_{\beta}I_{\alpha\beta}$.
Assuming that the probability distributions of fluctuating currents
$I_{\alpha\beta}$ are known, we
have to find an evolution of the distribution $\Gamma(Q,t)$
of the set of charges $Q= \{Q_{\alpha}\}$ for a given initial condition
$\Gamma(Q,0)$. In other words, one has to find 
an evolution operator $U(Q,Q',t)$ such that
$\Gamma(Q,t)=\int dQ'U(Q,Q',t)\Gamma(Q',0)$.
We use again the separation of time scales, follow the lines of the
derivation of Eq.\ (\ref{corr-S}), and obtain
the evolution operator $U=\exp[S_0(Q,Q',t)]$, 
with $S_0$ being a saddle-point
solution of the action
\begin{equation}
S=\int_{0}^{t} dt'[i
\chi\dot Q
+(1/2)\sum_{\alpha\beta}H_{\alpha\beta}
(\chi_{\alpha}-\chi_{\beta})],
\label{action1}
\end{equation}
where $\chi=\{\chi_{\alpha}\}$ is the set of charge counting fields. 
The generators $H_{\alpha\beta}$ of the 
cumulants of currents
$I_{\alpha\beta}$ have the obvious symmetry $H_{\alpha\beta}(\chi_{\alpha}
-\chi_{\beta})=H_{\beta\alpha}(\chi_{\beta}-\chi_{\alpha})$.

Next we assume that the ``Hamiltonian'' $H=(1/2)\sum_{\alpha\beta}H_{\alpha\beta}$ 
depends only on the subset of charges $Q^c=\{Q^c_{\alpha}\}$, 
which we call conserved charges 
since they are related to conserved quantities, such as total energy and charge. 
We complete the set $Q=\{Q^c,Q^a\}$ by the 
subset of non-conserved or absorbed charges $Q^a=\{Q^a_{\alpha}\}$. Since $H$ does not depend
on $Q^a$, we find that the corresponding ``momentums'' $\chi^a=\{\chi^a_{\alpha}\}$ 
are integrals of motion: $\chi^a=const$. This allows us 
to integrate out the corresponding terms in (\ref{action1}):
$i\int_{0}^{t} dt'\chi^a\dot Q^a=
i\chi^a[Q^a(t)-Q^a(0)]$. Thus the rest of the
action (\ref{action1}) becomes a generator of cumulants 
of the absorbed charge $Q^a(t)-Q^a(0)$.
In particular, the stationary solution for the cumulant generator
is $S_0=tH(\chi^a,\bar\chi^c,\bar Q^c)$, where $\bar\chi^c$
and $\bar Q^c$ are solutions of the equations 
$\partial H/\partial\chi^c=\partial H/\partial Q^c=0$.

In the example of the elastic transport considered above there are 
3 reservoirs: Two leads ($\alpha=1,3$) and one cavity ($\alpha=2$).  
The state of the cavity is described by only one charge $Q_C=Q_2$, the total
charge on the cavity, while $Q=(Q_3-Q_1)/2$ is the absorbed charge 
which is being counted. The counting fields are $\chi_C=\chi_2$, and 
$\chi/2=\chi_1=-\chi_3$.
Next we consider the case of inelastic transport, which requires
the introduction of a new generalized conserved charge $E_C$, the total 
energy of the cavity.

{\em Hot-electron regime.}
We consider the electron transport through a cavity
in the so-called hot-electron regime, i.e.\ when the electron-electron
scattering time is much shorter than the relaxation time
$\tau_{ee} \ll \tau_C$, electrons in the cavity
then relax to local thermal equilibrium before
escaping from the cavity. Their distribution is given by
a Fermi function $f(\varepsilon) = \{1+\exp[(\varepsilon-\mu_C)/T_C]\}^{-1}$
where $\mu_C$ and $T_C$ denote respectively the electro-chemical potential 
and electron temperature in the cavity.
The noise in the hot-electron regime was first considered
theoretically for diffusive wires \cite{Rudin1}.
This calculation and its experimental verification \cite{Henny1}
found that electron-electron interaction increases shot
noise. Similar results for the chaotic cavity were presented 
in Ref.\ \cite{Beenakker1} and measured in Ref.\ \cite{Oberholzer1}.
We now derive the FCS of transmitted charge $Q$.

Our calculation starts from the observation that the nonequilibrium state
of the cavity is fully described by only two parameters, $\mu_C$ and $T_C$.
They can be expressed in terms of conserved values: The total charge  
$Q_C = e^{-1}C_{\mu}\mu_C$ and the total energy 
$E_C = (1/2C_{\mu})Q_C^2+Q_CV_G +(\pi^2/6)n_FT_C^2$ of the cavity, 
where $C_{\mu}$ is the electrochemical capacitance of the cavity,
$1/C_{\mu}=1/C+1/(e^2n_F)$,
and $V_G$ is the gate voltage.
The charge and energy conservation can be written as
\begin{eqnarray}
&&\int d\varepsilon\,
[I_L(\varepsilon)+I_R(\varepsilon)]=-\dot Q_C,
\label{conserv-law1} \\
&&\int d\varepsilon\,\varepsilon\,[I_L(\varepsilon)+I_R(\varepsilon)]
=-e\dot E_C,
\label{conserv-law2}
\end{eqnarray}
where $I_{L,R}(\varepsilon)$ are the outgoing currents through the point contacts
per energy interval $d\varepsilon$. We now replace the constraint
(\ref{constraint}) with these two equations, and apply it to the 
energy resolved analogue of path integral (\ref{uncorr-S}). By doing so,
we introduce the counting field $\chi_E$ for the total energy $E_C$. 
Instead of (\ref{Hamiltonian}), the new ``Hamiltonian'' 
$H(\chi,\chi_C,\chi_E;\mu_C,T_C)$ takes the form
\begin{equation}
H  = \int d\varepsilon[ H_L(\chi_C+\varepsilon\chi_E-\chi/2) 
+H_R(\chi_C+\varepsilon\chi_E+\chi/2)],
\label{Hot Hamiltonian}
\end{equation}
with $H_{L,R}$ given by Eq.\ (\ref{H-LR}).

To obtain the FCS
in the long time limit $t\gg\tau_C$, where 
$\tau_C=2\pi\hbar C_{\mu}/[e^2(N_L+N_R)]$
is the RC-time (the time of charge screening in the cavity
\cite{Brouwer}), 
we need to evaluate  $H$ as a function of 
$\chi$ at the saddle point. We find the saddle-point solution $S_0(\chi,t)$
numerically and calculate $P(Q,t)$ by Fourier transformation
(\ref{P-def}). The result is shown in Fig.\ 
\ref{Probability Distribution}. It is clearly visible that noise is
enhanced in the hot-electron regime.

In order to make further analytical progress
we note that the saddle point solution for
$\chi=0$ is given by the chemical potential
$\mu_C = (N_L \mu_L + N_R \mu_R)/(N_L+N_R)$ and the effective temperature
$T_C = \sqrt{T^2 + T^2_V}$, where $T$ is the
temperature of the reservoirs, and
$T_V = (\sqrt{3}/\pi)\eta eV$ 
is proportional to bias with
$\eta^2 = N_LN_R/(N_L+N_R)^2$.
The internal fields $\chi_C,\chi_E$ are
zero for $\chi=0$. It is now
straightforward to expand Eq.
(\ref{Hot Hamiltonian}) and the corresponding
saddle point equations around $\chi = 0$
and to calculate higher cumulants order by order.
This procedure is very similar to the cascade
approach considered in Refs.\ \cite{Nagaev1,Nagaev2}. 
For the first few cumulants we obtain
\begin{subequations}
\label{hotcums}
\begin{eqnarray}
\langle Q^2\rangle 
&=&\frac{e^2\eta t}{2\pi\hbar}
\sqrt{N_LN_R}\;\left(T+T_C\right),
\label{hotcum2}\\
\langle Q^3\rangle 
&=& -\frac{e^3\eta^2t}{2\pi\hbar}
\sqrt{N_LN_R}\;\frac{3\sqrt{3}TT_V}{\pi T_C}\,,
\label{hotcum3}\\
\langle Q^4\rangle 
&=& \frac{e^4\eta^3t}{2\pi\hbar}\sqrt{N_LN_R}
  \frac{9}{\pi^2}\left[T-T_C + \frac{2T^4}{T_C^3}
\right], 
\label{hotcum4}
\end{eqnarray}
\end{subequations}
where the second cumulant coincides with the one in the
Ref.\ \cite{Oberholzer1}.
The result for the third cumulant shows that odd cumulants are
strongly suppressed. We note that screening does not affect 
the FCS in the long time limit $t\gg\tau_C$.

{\em Conclusions.}
We have constructed a stochastic path integral formulation
of full counting statistics. Our approach is based
on the fact that fast microscopic quantum fluctuations give rise to
slow variations of conserved quantities. The method has been illustrated
with chaotic cavities and novel results have been presented
for the hot-electron regime. We emphasize the general nature of
this approach and its applicability to
stochastic problems even outside mesoscopic physics.

S. P. acknowledges the collaboration with K. E. Nagaev on an earlier
related work. We thank P. Samuelsson for useful discussions. This work was
supported by the Swiss National Science Foundation.


\begin{thebibliography}{02}
\bibitem{commentm1}
We assume that the flight times through the contacts and cavity
are the shortest time scales.
\bibitem{comment1}
The currents $I_{L,R}=\dot Q_{L,R}$ are Markovian on
the time scale $\Delta t\gg\tau_0$, implying
the semigroup property
$P_l(Q_l,t_1+t_2)=\int dQ'_lP(Q_l-Q'_l,t_2)P(Q'_l,t_1)$, 
$l=L,R$.
\bibitem{comment2}
The integration in (\ref{corr-S}) goes over the whole set 
of variables $\{\chi_C(t_n),Q_C(t_n)\}$ except the initial 
condition $Q_C(0)$. According to Eq.\ (\ref{constraint}) $H$ does not depend 
on $Q_C(t_{N+1})$. Thus the integration over $Q_C(t_{N+1})$ in (\ref{corr-S}) fixes
the condition $\chi_C(t_N)=\chi_C(t)=0$.
\bibitem{SP-solution}
A similar result was found in Ref.\ \cite{Kindermann1}
using a fully microscopic theory based on Keldysh technique.
\bibitem{comment3}
This is true up to a boundary term $\int dt'Q_C\dot Q_C$.
\bibitem{comment4}
The contribution $S_{fl}$ depends on the approximate time discretization procedure 
(\ref{constraint}), and thus the numerical prefactor in $S_{fl}$ is beyond our formalism
(this is related to the usual operator ordering ambiguities in path integrals
\cite{Langevin}). Fortunately, as we have shown, $S_{fl}$  is small and can be neglected compared $S_0$.
\end{thebibliography}
\end{document}